\documentclass[    final     ]  {aipproc}

\layoutstyle{6x9}

\begin{document}

\title{3D ISM-Shock Spectral Emission: X-ray  models for  Radio Galaxy SED Modeling} 

\author{Ralph  S. Sutherland}{
  address={Research School of Astronomy and Astrophysics, \\ 
  Australian National University, ACT 0200, Australia}
}

\begin{abstract}
Galaxies form out of small fluctuations in a smoothly expanding
Universe.  However, the  initial gravitational collapse phase  is
accompanied by the formation of supermassive black holes and clusters
of massive stars.   Black holes and star clusters generate powerful
outflows in the form of jets and superwinds that interact with
still infalling gas, possibly regulating the galaxy formation
process, initiating new sites of star formation, and carrying
chemically enriched gas to the intergalactic medium. Unfortunately,
beyond this qualitative description our detailed
theoretical understanding is poor.

New results from 3D simulations of a GPS/CSS galaxy,
with gravitational potentials included, shed some new light on
the jet driven outflow process in particular. New code  capabilites to 
predict detailed X-ray spectra from multi-dimensional time-dependent 
dynamics simulations of Galaxy Feedback, and will be useful for future
interpretation of X-ray and radio  SEDs of forming galaxies.

\end{abstract}

\maketitle

%%%%%%%%%%%%%%%%%%%%%%%%%%%%%%%%%%%%%%%%%%%%
%% MAINMATTER
%%%%%%%%%%%%%%%%%%%%%%%%%%%%%%%%%%%%%%%%%%%%

\section{Introduction}

Key questions about the formation and evolution of galaxies over the history of the Universe include :

\begin{itemize}
\item {\bf Intergalactic medium enrichment:}  How is the IGM enriched in the early universe to provide 
the non-zero metallicity  material that formed the observed population II stars in galaxies like our own. 
What role do energetic outflows play in carrying metals out of large potentials, or are outflows in small halos more important? 

\item {\bf Halo baryon mass fractions}  Measurements of baryon mass fractions in present day halos show a deficit 
compared to the inferred universe average.  What role does  galactic formation feedback play in bringing this about, and how are baryons removed from the most massive halos where starformation alone may be insufficiently powerful.
(e.g. \citet{silk04a}).

\item {\bf The black hole halo correlation} The observed correlation between the central black hole mass and the holst halo velocity dispersion indicates the action of a strong coupling between the blackhole growth and the star formation in the host halo.  What this connection is remains open to debate. (e.g.  \citet{magorrian98a,tremaine02a}).

\item {\bf High Redshift Radio Galaxies} What can early galaxies at high redshift tell us about starbursts, hot halos and central blackhole activity.  What models do we need to interpret the observations. (e.g., \citet{adelberger03}).

\end{itemize}

Panchromatic (very broad range) SEDs may contain indirect and often subtle information about many of these physical problems -- and we require broad band detailed models to interpret the SEDs fully.

\section{Theoretical Approach: ISM Thermal Emission}

One approach to understanding these issues is to focus on one particular component of the physical processes that contribute to the SEDs, so that that component's contribution can be characterized and extracted from an observed SED.  In these proceedings a lot of effort is has been put into detailed models of the radiative transfer through dusty media to help understand the optical and infrared spectra of galaxies in differnt stages of evolution.  Here instead I will focus on the atomic and ionized interstellar medium plasma (ISM) with an emphasis on the X-ray emission.

The behavior and properties of the observed ISM plasma can inform us about the underlying galaxian processes, and the X-rays are often penetrating enough to escape shrouded regions  where otherwise only IR radiation may be expected to escape: 

From observations of galactic  ISM  we might hope to learn about :
\begin{itemize}
\item {\bf The input of energy and momentum:} such as  galactic winds,  radio jets
\item {\bf The distribution of energy sources:} starbursts,  clusters, AGN
\item {\bf Composition from plasma diagnostics:} gas enrichment histories.
\end{itemize}

However before we can get quantitative estimates of any of the above, which are important if we are to learn answers to the key questions, we will require detailed a ISM model.  This will have to include: Microphysics- (ionisation,excitation,  molecular chemistry, dust physics, radiative transfer),
Excitation Mechanisms: (shocks, photoionisation nebulae) as well as the ability to accomodate 
phase structure and distributions of the ISM and its dynamical radiative properties.

\subsection{Hypersonic Shockwaves}

Hypersonic (Mach $> 5$) shocks are a ubiquitous and efficient means of transforming kinetic energy into hot thermal plasmas and emission.  In the ISM they are common, and are driven by starformation through stellar jets, wind bubbles and supernova explosions. In AGN as well a central black hole powered engine can provide powerful shocks through jets and radiation driven outflows.  This generic ISM process is a good candidate for a component of a detailed ISM model for spectral synthesis.

\subsubsection{A 3D X-ray Spectral Synthesis Shock Model}

\begin{figure}
 \includegraphics[height=0.5\textheight]{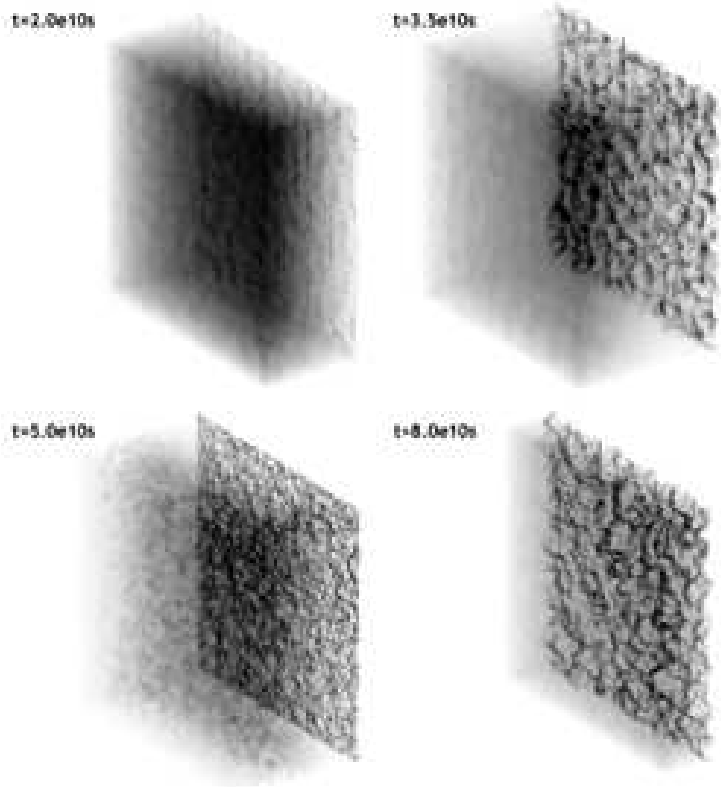}
  \caption{\footnotesize Very soft X-ray (200-500 eV) emission from a 200 km/s inhomogeneous interstellar medium 3D shock, in an ambient medium of density $\sim 1$ H atom cm$^{-3}$   with 10\% R.M.S. density fluctuations.  Upper left: Initially the shock is relatively smooth and uniform in temperature, and grows steadily into the incoming flow of gas (flow from left to right, shock travelling to the left).  Upper Right: the centre of the shocked layer cools, reducing the X-ray emission in the cenre, and forward progress halts.  Lower Left: the shock collapses, with local density contrasts increasing at the same time, giving a filamentary structure in the collapsing gas.  Lower Right, eventually a shock reforms and the process repeats.  The dense layer on the right becomes more turbulent and distorted, and some of the model X-rays from this region are over-estimated in due to numerical diffusion.  The simulation was carried out at $768^3$ celll resolution ($\sim 452\;$million cells).  Each cell produces an 1100 frequency X-ray spectrum, which are filtered and co-added to make these optically thin images.}
\label{fig:shock}
\end{figure}

An important aspect of the dynamical modelling pesented here is the emphasis on producing results that are closely aligned to the observational plane, to allow for more direct comparison with data such as X-ray and radio telescope images.  A  3D hydrodynamical code, \emph{ppmlr}, with radiative emission is used, to which X-ray spectral synthesis has been added. Figure  \ref{fig:shock} shows an example model of a Mach $\sim 20$ shockwave of about 200 km/s on average with the new code.  The thermal cooling in the model produces both local thermal instability, enhancing local density contrasts during cooling, as well as global shock pulsations.

\begin{figure}
 \includegraphics[height=.25\textheight]{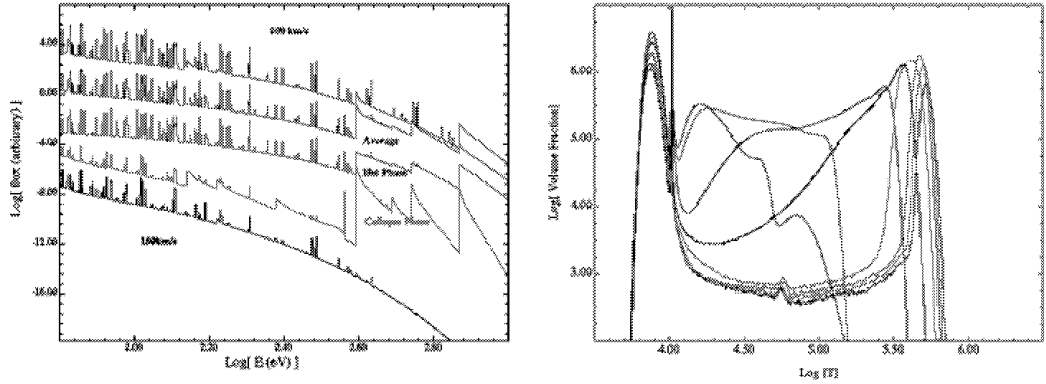}
  \caption{\footnotesize Dynamical Shock Spectra and Thermal Distributions.  Left: the instantaneous 3D shock X-ray spectra at different stages of evolution, compared with two steady 1D shock models at 150 and 200 km s$^{-1}$.  Right: the instantaneous thermal distribution of gas in the shocked region, by volume, showing the changes in distribution over time.  The spike at $\log[ T ] = 4.0$ is the contribution from the pre--shcok and post--shock photoionised zones. }
\label{fig:sspec}
\end{figure}

The new modelling code now predicts the optically thin X-ray emission from the shock models, looking up the spectra associated with each point on the cooling function calcluated by the MAPPINGS III code (version m).  Figure \ref{fig:sspec} shows the integrated spectra from 100eV to 1keV for the model at various times in an expansion/collapse cycle.  These are compared to two steady shock spectra from MAPPINGS.  The dynamical spectrum, including an ensemble average of all the spectra, cannot be approximated in detail with the steady models. Different regions of the spectra are dominated non-linearly by shocked gasses of different temperatures so the average does not match any one steady model.  The dimensional dependence of turbulent structures, and the non-linear  ensemble averages of time dependent unsteady shocks both prevent a steady shock approximation in detail.   This is one reason for pursuing fully 3D global radiative hydrodynamical simulations, in preference to fitting steady models to components of  a composite model.  Modern computing techniques and hardware are now capable of computing $0.1-1.0$~billion cell models giving the spatial range required for the first time.

This 3D model will be analysed and presented in detail in a future publication.  These 3D models will give new spectra for the ionising field produced by shockwaves in the ISM.  The main limitations of new models presented here are that they avoid, ( for now ), the difficult 3D radiative transfer problem, restricting results to optically thin X-ray models and radio emission, and they are a compromise between complete self-consistency and speed of execution in that the spectra for each cell are generated by lookup tables for each temperature, albeit from detailed 1D non-equilibrium cooling calculations.  

\section{Dynamic Active Galaxy X-ray SED}

We now present a new, initially qualitaitve, 3D active radio galaxy simulation, using the shock physics (partially resolved for slow shocks) outlined above, but here in a global simulation with complex geometry.  Many of the shocks will be poorly resolved, however the major structures should be captured.
Additional physics is added to the simulation, principlaly the gravitational  potential  of the host galaxy, a high velocity radio jet of non-thermal plasma, and powerlaw clumpy ISM with a high contrast, log-normal density distribution to better account for the geometry of the ISM in a way that a simple smooth model cannot.

\subsection{Dynamic Galaxy: Host Gravity and ISM}

The gravitational potential of the model host galaxy was approximated as a combination of a dark matter halo (Potential $\Phi_H$, mass = $ M_{H} = 10^{12}$M$_\odot$, Core radius $= 500\;$pc) and a stellar spheroid  (Potential $\Phi_{\rm ss}$, mass = $M_{H} = 10^{11}$M$_\odot$, Core radius $= 350\;$pc).   As this initial investigation is primarily qualitative, two simple potential approximations were used.   Future simulations will use more detailed potentials.  An  empirical Navarro, Frenk \& White (\citet{navarro1997})  widely seen in numerical n-body simulations was used for the dark matter, while an approximation to the truncated King isothermal profile (ref) was used for the stellar spheroid: The Navarro--Frenk--White Empirical Potential was written as,
$ w  =  r/r_{s}\; \; , \Phi_s(r)  =  - {GM_s}/{r} \; \log\left(1 + w\right)\, $,
 and the Truncated King Potential Approximation (\citet{king1966}) as,
$ u  =  r/r_{ss}\; \; , \Phi_{ss}(r)  =  - {GM_{ss}}/{r} \; \log\left(u + \sqrt{1 + u^2}\right) \, .$
In each case the domain of the simulation is within a few  core radii. 
With these analytic potentials, a hydrostatic atmosphere was constructed with a mean pressure $p/k = 10^6$ and a combination of thermal phases, using $\rho / \rho_0  = \exp ( \sigma_{\rm eff} ^2 \Phi)
$, where $\sigma^2$ is the effective velocity dispersion (or alternatively effective gas temperature $T_{\rm eff} = \mu m \sigma^2/k$), and $\rho$ is the gas density.  

A smooth hot gaseous halo with $T_{\rm eff} = 10^7\;$K was placed in hydrostatic balance with the total potential $\Phi = \Phi_H + \Phi_{\rm ss}$.  Then, a component was added with a much lower effective temperature (assumed to be a combination of intrinsic thermal temperature and turbulent support), using a two point statistical power law fractal distribution with a Kolmogarov density power spectrum $P(k) \propto k^{-5/3} = k^2 \Psi(k)$,  where $\Psi(k) \propto k^{-11/3}$ is the 3D fourier spectral energy density, and single point log-normal density statistics, $P(\rho) d\rho= 1/({s \sqrt{2 \pi} \rho})\, {\exp[ -(\ln(\rho)-m)^2/2s^2 ]}\, d\rho$.  This represents the multiphase host ISM,  the atomic, ionized and molecular phases.   

 The log-normal parameters $m$ and $s$ are related to the mean, $\mu$, and the variance, $\sigma^2$, of the density field by $m = \ln[\mu^2/\sqrt{\sigma^2 + \mu^2}]$, and $s^2 =  \ln[(\sigma^2/\mu^2) +1]$.  A density field with $\mu = 1.0 $ and  $\sigma^2 = 5.0$ was created via 3D Fourier transforms, and then scaled the density as required by hydrostatic equilibrium.  The effective temperature for this component was taken as $50,000\;$K, with a resulting mean density of $20\;$cm$^{-3}$, giving a  median and modal densities of 8.165 and $1.361\;$cm$^{-3}$ respectively.  

\section{Results}

\begin{figure}[th!]
 \includegraphics[width=0.7\textwidth]{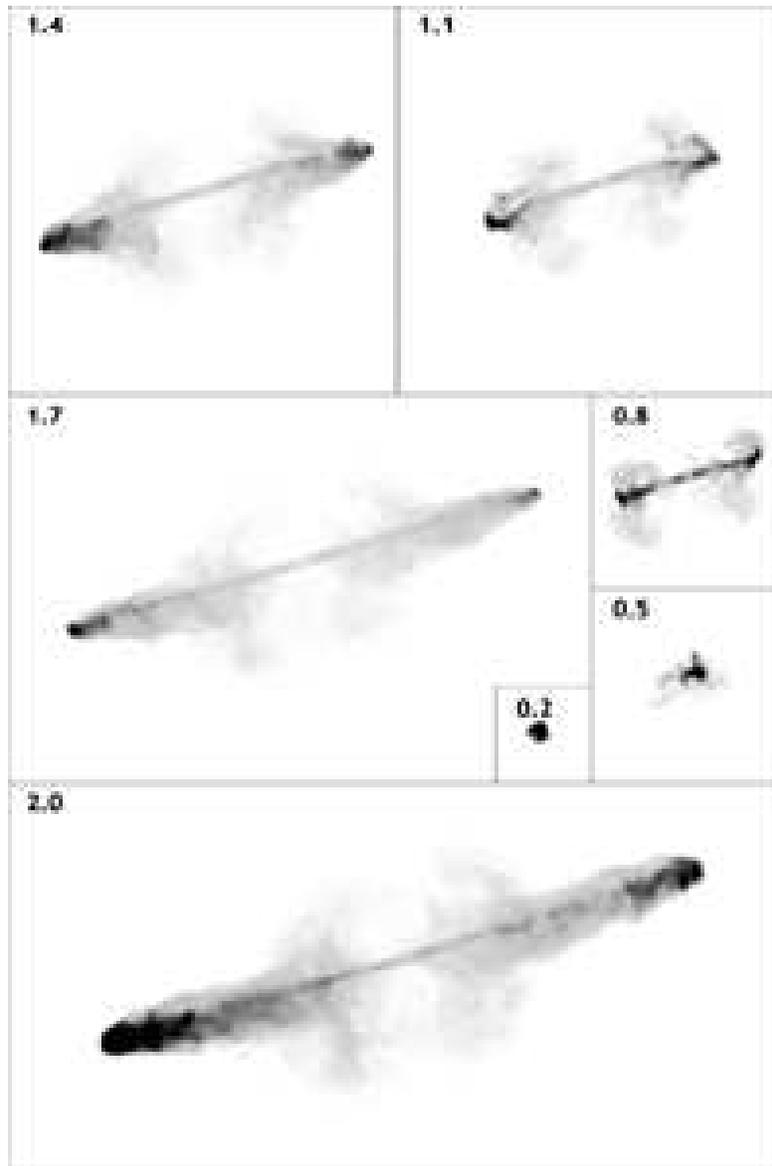}
  \caption{\footnotesize Radio Evolution.  Spiraling out anti-clockwise, the frames show radio emission maps, assuming total emission$\propto p^{1.8}$ in the non-thermal plasma, and transforming as the square root of the resulting surface brightness, as is commonly used in presenting radio images.}
\label{fig:radevolve}
\end{figure}

\begin{figure}[th!]
 \includegraphics[width=.7\textwidth]{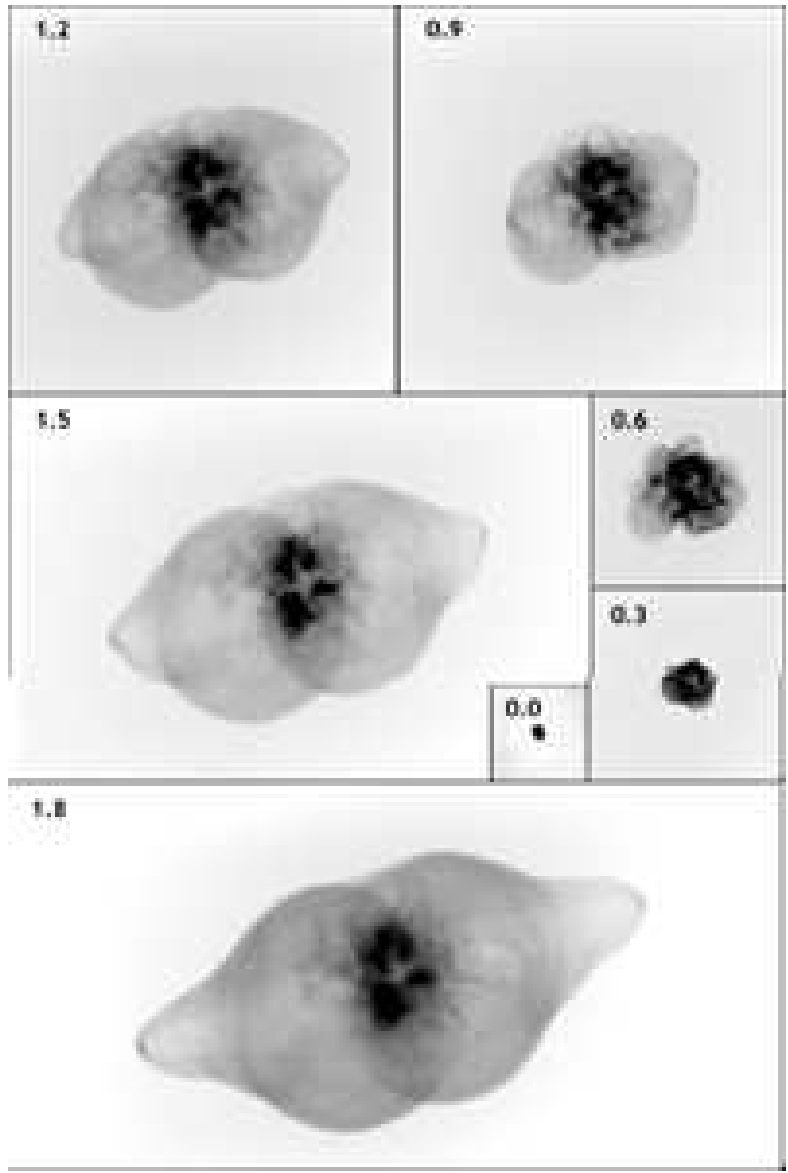}
  \caption{\footnotesize X-ray Evolution. Spiraling out anti-clockwise, the frames show x-ray emission maps, energy rnage from 0.1-10 keV.  Time is indicated in relative model units, so the last frame ($t=2.0$) is 4 times older than the second frame at $t= 0.5$. The simulation was performed on one side only, these double sided structures were created by mirroring the $x$-axis.}
\label{fig:xevolve}
\end{figure}

A simulation was run with a resolution of $256^3$ cells, set initially to cover a domain of $0<x<1000\;$pc, $-500<y<500\;$pc and $-500<y<500\;$pc.  The cell elements then represented a volume of $4\times 4\times 4\;$pc.  Sequences showing the evolution of the pressure in regions occupied by the jet plasma, traced by a scalar tracer variable, and the thermal X-ray emission from 1.0 to 10.0 keV are shown in figures \ref{fig:radevolve} and  \ref{fig:xevolve}.  These images were rendered in a post-processing code, which added the second side of the radio source by mirror symmetry along the $x-$axis.

\section{Interpretation}

The interpretation of this sequence of models remains highly preliminary.  However we can identify a number of key elements that are currently under investigation in more detail.

Firstly, the behavior of the system is initially one where the dynamics are dominated by the escape of the high pressure non-thermal plasma from the clumpy medium in the centre of the galaxy.  Once the jet finds a channel it begins to blow an energy driven bubble that is more or less spherical.  The bulk of the clumpy ISM is then shocked as the high pressure bubble created by the jet expands and a primary shock crosses the plane of the galaxy.    Rather than simply sweep out the ISM in a uniform shock in a few crossing time scales, the dense cores of the clumps remain in place, and only the outer low density regions are heated sufficiently to expand and be swept up.    Very little of the clumpy ISM interacts directly with the jet. ($t = 0.0$ to $\approx 0.7$ in figures \ref{fig:radevolve} and  \ref{fig:xevolve} ).

Then follows a short phase where the initially disrupted jet straightens up and begins to establish
endpoint Mach-disks.    ($t = 0.6$ to $\approx 0.9$ in figures \ref{fig:radevolve} and  \ref{fig:xevolve} ).
Once a clear channel has established, the jet stabilises in the model, and the large scale vortexes of previous 2D models (e.g. \citet{bicknell03a,bicknell03b}) are not present, reducing feedback interference of the jet by the cocoon gas. The hot thermal bubble is still quasi-spherical at this point.

Finally the straightened jet moves forward into the outer atmosphere where the density and pressure are decreasing in a power law fashion.  We speculate that as the jet becomes heavy compared to the ambient density, it eventually becomes heavy jet, and proceeds to drill into the outer halo in a momentum driven mode.  This produces the characteristic extensions on the thermal bubbles, and correponds to the development of the classical FRII radio lobes ($t  > 1.2$ in figures \ref{fig:radevolve} and  \ref{fig:xevolve} ).

\begin{figure}[th!]
\includegraphics[width=.7\textwidth]{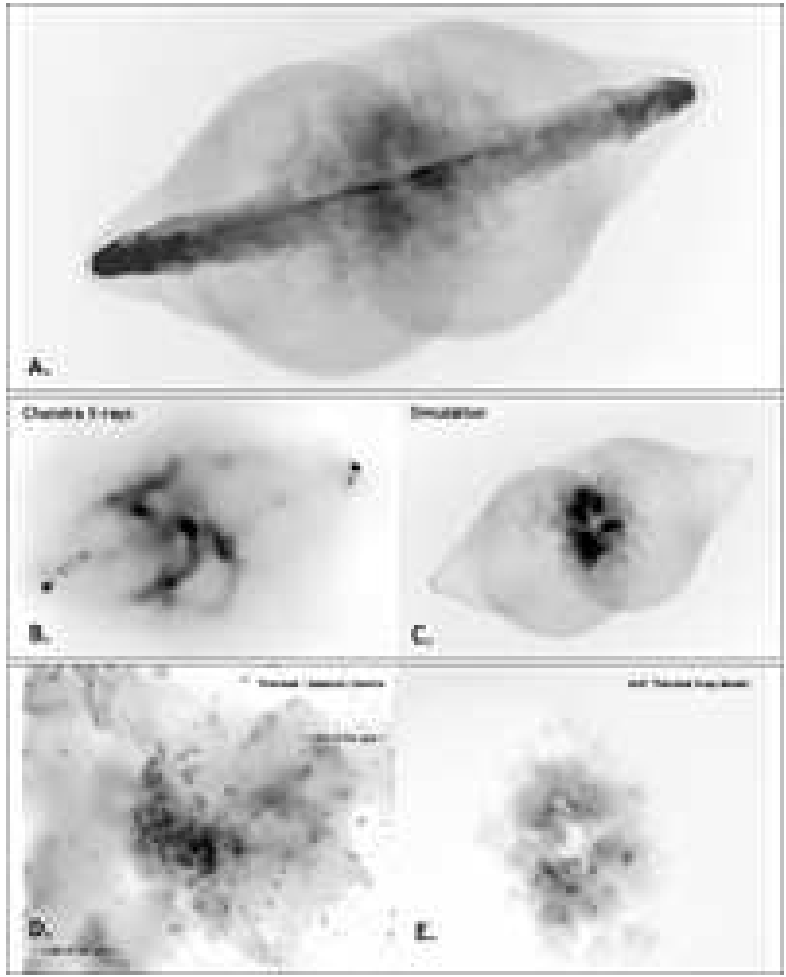}
\caption{\footnotesize {\bf A.},  Radio Jet  -- X-ray overlay.  This composite image shows how the radio plasma fits inside the X-ray bubble structure.  The 'skirt' of the radio emission is associated with the broad parts of the bubbles and are related to the original halo atmosphere structure, being inside a radius that was dense compared to the jet density. {\bf B.},   Cygnus A X-ray comparision.  The adaptively smoothed 1-10 keV chandra image of Cygnus A from \citet{smith2002}.  {\bf C.},  the broad band X-ray emission of the model at approximately $t = 1.6$ in model units, oriented to match the apparent projection of Cygnus A.  At least some of the X-ray features in the observations may be projection effects from the overlap of the double bubble structure, others are cooler and may be ablated ISM material from the host or in-falling satellites in the past. {\bf D.},  Comparison of  X-rays with Galactic Centre.  The Chandra X-ray image by \citet{baganoff2003}, \citet{muno2003}.  {\bf E.},  the simulation rendered in  soft X-ray bands (0.2 - 1.2 keV) that show only the low temperature shocked ISM material.  Each show dense cloud cores, embedded in  an intermediate density medium of gas that may be ablated from the cores by non-radiative shocks in the less dense parts of the original clumpy medium.  }
 \label{fig:comps}
\end{figure}

Despite the simulation parameters being set for a relatively small halo and moderate jet power, there is a  striking similarlity of the morphology of both the radio and thermal emission of the model in the later stages (  $t = 1.4$ to $\approx 2.0$ in figures \ref{fig:radevolve} and  \ref{fig:xevolve} ) to the nearby quasar  3C405 (Cygnus A, a far from exhaustive list of key references from the vast Cygnus A literature include: \citet{boulton1948, perley1984,  carilli1996, wilman2000, smith2002, taylor2003}).  This nearby source ($z = 0.0562$, and $1" = 1.003\; $kpc, based on $H_0 = 75\; $km/s/Mpc, $q = 0$. \citep{smith2002}) is a cD Ellipitcal with a clumpy ionised and molecular ISM in the inner regions, plus a $>10^{46}$ erg s$^{-1}$ jet.  It has been shown to exhibit a bi-polar thermal bubble in 1-10 keV X-rays with a major axis of $\sim 62$kpc , as well as the typical FRII lobed powerful radio galaxy morphology.  The radio plasma fits inside the X-ray cocoon, as shown in figure \ref{fig:comps}.{\bf A}, at $t \approx 1.9$.

The correspondence between the model and Cygnis A in thermal X-rays \citet{smith2002}  shown in figure \ref{fig:comps}. {\bf B-C.} is striking.   This correspondence suggests a degree a scalability in the models, which warrants further investigation.

Finally we note that the clumpy ISM survives for a very long time in the simulation, as ever decreasing shock velocities slowly traverse ever increasing clump densities.  After many shock crossing times for the main blast wave there remain bright cores of soft X-ray emission, as shown in figure \ref{fig:comps}.{\bf D-E}.  

A speculation is that it is the fractal nature of the ISM and the properties of the skewed log-normal density distributions that contributes to this robustness.  No matter what ram pressure the ISM is hit with, there will always be a density present along the density distribution where the induced cloud shocks are radiative and serve to further compact the cores - making them resistant to ablation.  Below this density, the induced shocks will be non-radiative on the interaction time scale and will heat the gas, which in turn will be advected in the bubble expansion.  The single point density statistics (the log-normal distribution) will determine the volume and mass fraction that will be advected easily, while the two point statistics (the fractal power spectrum or structure functions) will affect the efficiency by determining the connectivity or otherwise of the threshold density regions and how the main blast can pass by or push on the clumps.   The study of these advection factors forms the basis of ongoing research, and will be reported on in future works.  For the present we note that the presence of clumped ISM, and even molecular gas in the inner regions of the Galactic Center and AGN such as Cygnus A may be ubiquitous, and not all be due to an organized torus, but could be in a clumpy ISM that is resistant to being swept out owing to its non-geometric and non-homogenous distribution.

\section{Conclusions}

A new experimental model of a radio jet outflow inside a galactic halo + dark matter potential shows a
dynamical sequence, inside and then outside the core radius of the galaxy. The X-ray SED of the model as it evolves shows signs of the interaction of the radio jet and bubble with the dense clumpy host ISM in the soft X-rays, and the interaction with the hot halo in the hard X-rays.  The soft X-rays persist to late times, possibly due to the clumpy nature of the dense ISM that allows cloud cores to survive for many crossing times of the energy driven blast wave powered by the jet.   The radio morphology is suggestive of an powerful FRII radio source and matches the morphology of Cygnus A on many levels, even though the model was set up to simulate a low powered GPS source.  This implies that overall the large scale  processes are relatively scale free, even though some small scale ones clearly are not.  This will be investigated in future modeling.

%%%%%%%%%%%%%%%%%%%%%%%%%%%%%%%%%%%%%%%%%%%%%%%%
%% BACKMATTER
%%%%%%%%%%%%%%%%%%%%%%%%%%%%%%%%%%%%%%%%%%%%%%%%

\begin{theacknowledgments}
Thanks are due to Dr  Geoff Bicknell for many insightful discussions on the experimental simulations presented here, and to the ANU supercomputing facility/APAC for the computer time allocations.
This research and the travel was funded by an ARC discovery project grant DP0208445.
\end{theacknowledgments}

%% If the bibliography is produced without BibTeX comment out the

\bibliographystyle{aipproc}   % if natbib is available

\bibliography{suth04}

\IfFileExists{\jobname.bbl}{}
 {\typeout{}
  \typeout{******************************************}
  \typeout{** Please run "bibtex \jobname" to optain}
  \typeout{** the bibliography and then re-run LaTeX}
  \typeout{** twice to fix the references!}
  \typeout{******************************************}
  \typeout{}
 }

\end{document}